\documentclass[prd,onecolumn,superscriptaddress,nofootinbib,showpacs,showkeys]{revtex4}
\usepackage{epsfig}
\usepackage{amssymb}
\usepackage{amsmath}
\usepackage{dcolumn}

\newcommand{\NIMA}[3] {Nucl.\ Instr.\ and Meth.\ \textbf{A#1} (#2) #3}

\newcommand{\Zzero}{\mbox{${\mathrm{Z}}$}}

\newcommand{\qq}{\mbox{$\mathrm{q\overline{q}}$}}
\newcommand{\uu}{\mbox{$\mathrm{u\overline{u}}$}}
\newcommand{\qqp}{\mbox{$\mathrm{q'\overline{q}}$}}

\newcommand{\EGeV}{\mbox{$E(\mathrm{GeV})$}}

\newcommand{\thetaqq}{\mbox{$\theta_{\mathrm{q\overline{q}}}$}}
\newcommand{\cosqq}{\mbox{$\cos\thetaqq$}}

\newcommand{\GeV}{\mbox{$\mathrm{GeV}$}}

\newcommand{\roots}{\mbox{$\sqrt{s}$}}

\newcommand{\rms}{\mbox{${\mathrm{rms}}_{90}$}}

\newcommand{\MARLIN}{\mbox{\sc Marlin}}
\newcommand{\MOKKA}{\mbox{\sc Mokka}}
\newcommand{\PANDORAPFA}{\mbox{\sc PandoraPFA}}
\newcommand{\GEANT}{\mbox{GEANT4}}

\newcommand{\LCIO}{\mbox{\sc Lcio}}

\newcommand{\Cambridge}{Dept. of Physics, Cavendish Laboratory, Univ. of Cambridge, JJ Thomson Av., Cambridge CB3 0HE, UK}
\def\etal{\mbox{{\it et al.}}}

\begin{document}

\begin{flushright}
CU-HEP-06/17 
\end{flushright}

\title{Particle Flow Calorimetry at the ILC\footnote{To appear in Proceedings of LCWS06, Bangalore, India, March 2006.}
}

\date{\today}
\author{M.~A.~Thomson}
\affiliation{\Cambridge}

\keywords{calorimetry, particle flow}
\pacs{07.05.Kf, 29.40.Vj, 29.85.+c}
\begin{abstract}
One of the most important requirements for a detector at the
ILC is good jet energy resolution. It is widely believed 
that the particle flow approach to calorimetry is the key to 
achieving the goal of $0.3/\sqrt{\EGeV}$. This paper describes the current 
performance of the \PANDORAPFA\ particle flow algorithm. 
For 45\,GeV jets in the Tesla TDR detector concept, the ILC jet energy resolution goal 
is reached. At higher energies the jet energy resolution becomes worse and can be described
by the empirical expression: $\sigma_E/E \approx 0.265/\sqrt{\EGeV} + 1.2\times10^{-4}\EGeV$. 
\end{abstract}

\maketitle

\section{Introduction}

Many of the interesting physics processes at the ILC will be characterised by
multi-jet final states, often accompanied by charged leptons and/or missing transverse 
energy associated with neutrinos or the lightest super-symmetric particles. 
The reconstruction of the invariant masses of two or more jets will provide a powerful tool 
for event reconstruction and identification. Unlike at LEP, where kinematic
fitting\cite{bib:mwfit} enabled precise jet-jet invariant mass reconstruction 
almost independent of the jet energy resolution, at the ILC this mass reconstruction will 
rely on the detector having excellent jet energy resolution. The ILC goal is to achieve a 
mass resolution for $\mathrm{W}\rightarrow\qqp$ and $\mathrm{Z}\rightarrow\qq$ decays 
which is comparable to their natural widths, i.e. $\sim$2\,GeV. A jet energy resolution 
of $\sigma_E/E = \alpha/\sqrt{\EGeV}$ leads to a di-jet mass resolution of roughly 
$\sigma_m/m = \alpha/\sqrt{E_{jj}\mathrm{(GeV)}}$, where $E_{jj}$ is the energy of the 
di-jet system. At the ILC typical di-jet energies will be in the range $150-350$\,GeV, 
suggesting the goal of $\sigma_E/E = 0.3/\sqrt{\EGeV}$. This is more than a factor 
two better than the best jet energy resolution achieved at LEP, 
$\sigma_E/E = 0.6(1+|\cos\theta|)/\sqrt{E(\GeV)}$~\cite{bib:Aleph-jet}. 
Meeting the jet energy resolution goal is a major factor in the overall 
design of a detector for the ILC.

\section{The Particle Flow Approach to Calorimetry}

It is widely believed that the most promising strategy for achieving a jet energy 
resolution of $\sigma_E/E = 0.30/\sqrt{\EGeV}$ at the ILC is the particle flow 
analysis (PFA) approach to calorimetry. In contrast to a purely 
calorimetric measurement, particle flow requires the reconstruction of the four-vectors of 
all visible particles in an event. The reconstructed jet energy is the sum of the energies
of the individual particles. The momenta of charged particles are measured in
the tracking detectors, while the energy measurements for photons and neutral hadrons 
is performed with the calorimetric system. The crucial step of the particle flow algorithm is 
to assign the correct calorimeter hits to reconstructed particles, requiring 
efficient separation of nearby showers. 

Measurements of jet fragmentation at LEP have provided detailed information on the 
particle composition of jets (e.g.~\cite{bib:Knowles,bib:Green}).  On average, after the 
decay of short-lived particles, roughly 62\% of the energy of jets is carried by 
charged particles (mainly hadrons), around 27\% by photons, about 10\% by long-lived neutral 
hadrons ({\em e.g.} n/K$^0_{\mathrm{L}}$), and around 1.5\% by neutrinos.
Assuming calorimeter resolutions of $\sigma_E/E = 0.15/\sqrt{\EGeV}$
for photons and $\sigma_E/E = 0.55\sqrt{\EGeV}$ for hadrons, a jet energy resolution of 
$0.19/\sqrt{\EGeV}$ is obtained with the contributions from tracks, photons and neutral
hadrons shown in Tab.~\ref{tab:res}. In
practice it is not possible to reach this level of performance for two main reasons.
Firstly, particles travelling at small angles to the beam axis will not be detected.
Secondly, and more importantly, it is not possible to perfectly associate all energy 
deposits with the correct particles. For example, if a photon is not resolved from a 
charged hadron shower, the photon energy is not counted. Similarly, if some of the energy 
from a charged hadron is identified as a separate cluster the energy is effectively double-counted. 
This {\em confusion} degrades particle flow performance. 
The crucial aspect of particle flow is the ability to correctly assign calorimeter
energy deposits to the correct reconstructed particles. This places stringent requirements 
on the granularity of electromagnetic and hadron calorimeters. Consequently, 
particle flow performance is one of the main factors driving the overall ILC detector 
design. It should be noted that the jet energy resolution obtained for a 
particular detector concept is the combination of the intrinsic detector
performance and the performance of the PFA software.

\begin{table*}[htb]
\centering
\renewcommand{\arraystretch}{1.2}
\begin{tabular}{lcccc}
{Component}          & {Detector} & {Energy Fraction} & {Energy Res.} & {Jet Energy Res.} \\ \hline
Charged Particles ($X^\pm$) & Tracker           & $\sim0.6\,E_{\mathrm{jet}}$ & $10^{-4}\,E^2_{X^\pm}$  & 
   $<3.6\times10^{-5}\,E^2_{\mathrm{jet}}$ \\
Photons $(\gamma)$          & ECAL              & $\sim0.3\,E_{\mathrm{jet}}$ & $0.15\,\sqrt{E_\gamma}$ & 
   $0.08\,\sqrt{E_{\mathrm{jet}}}$ \\
Neutral Hadrons $(h^0)$     & HCAL              & $\sim0.1\,E_{\mathrm{jet}}$ & $0.55\,\sqrt{E_{h^0}}$  &
$0.17\,\sqrt{E_{\mathrm{jet}}}$ \\ 
\end{tabular}
\caption{Contributions from the different particle components to the jet-energy resolution
(all energies in GeV). The table lists the approximate fractions of charged particles, 
photons and neutral hadrons in a jet and the assumed single particle energy resolution. \label{tab:res}}
\renewcommand{\arraystretch}{1.0}
\end{table*}

\section{The PandoraPFA Particle Flow Algorithm}

\PANDORAPFA\cite{bib:pandorapfa} is a {C++} implementation of a PFA algorithm running in the 
\MARLIN\cite{bib:marlin,bib:wendt} framework.
It was designed to be sufficiently generic for ILC detector optimisation studies and
was developed and optimised using events generated with the \MOKKA\cite{bib:mokka} program, which
provides a \GEANT\cite{bib:geant4} simulation of the 
Tesla TDR\cite{bib:teslatdr} detector concept. The \PANDORAPFA\ algorithm
performs both calorimeter clustering and particle flow in a single stage. The 
algorithm has six main stages: 

\noindent{\bf{i) Tracking:}} for the studies presented in this paper, the track pattern recognition is
               performed using Monte Carlo information\cite{bib:marlin}. The track parameters are then 
               extracted using a helical fit. The projections of tracks onto the 
               front face of the electromagnetic calorimeter are calculated using helical 
               fits (which do not take into account energy loss along the track).
               Neutral particle decays resulting in two charged particle tracks ($V^0$s) are 
               identified by searching for pairs of tracks which do not originate from the interaction 
               point and that are consistent with coming from a single point in space. 
               Kinked tracks from charged particle decays to a single charged particle and a 
               number of neutrals are also identified. When a kink is identified the parent 
               track is usually removed for the purposes of forming the reconstructed particles.

\smallskip
\noindent
{\bf{ii) Calorimeter Hit Selection and Ordering:}} isolated hits, defined on the basis of proximity 
          to other hits, are removed from the initial clustering stage. 
          The remaining hits are ordered into {\it pseudo-layers} which follow the detector 
          geometry so that particles propagating outward from the interaction region will 
          cross successive pseudo-layers. In most of the calorimeter the pseudo-layers 
          follow the physical layers of the calorimeters except in the barrel-endcap overlap 
          region and where the ECAL stave structure\cite{bib:teslatdr} 
          results in low numbered layers which are 
          far from the front face of the calorimeter. The assignment of hits to 
          pseudo-layers removes the dependence of the algorithm on the explicit detector 
          geometry whilst following the actual geometry as closely as possible. 
          Within each pseudo-layer hits are ordered by decreasing energy.

\smallskip
\noindent
{\bf{iii) Clustering:}} the main clustering algorithm is a forward projective method working 
                 from innermost to outermost pseudo-layer. In this manner hits are added 
                 to clusters or are used to seed new clusters. Throughout the 
                 clustering algorithm clusters are assigned a direction (or directions) 
                 in which they are growing. The algorithm starts by {\em seeding} clusters
                 using the projections of reconstructed tracks onto the front face of the
                 calorimeter. The initial direction of a track-seeded cluster is obtained 
                 from the track direction. The hits in each subsequent pseudo-layer are 
                  then looped over. Each hit, $i$, is compared to each clustered hit, $j$, 
                  in the previous layer. The vector displacement, ${\bf r}_{ij}$, is calculated and 
                  is used to calculate the parallel and perpendicular displacement of the 
                  hit with respect to the unit vector(s) ${\bf\hat{u}}$ decribing the
                  cluster propagation 
                  direction(s), $d_\parallel = {{\bf r}_{ij} .{\bf\hat{u}}}$  and 
                  $d_\perp = |{\bf{r}}_{ij}{\bf{\times\hat{u}}}|$. 
                 Associations are made using a cone-cut, 
                 $d_\perp < d_\parallel\tan\alpha + \beta D_{\mathrm{pad}}$, 
                 where $\alpha$ is the cone 
                 half-angle, $D_{\mathrm{pad}}$ is the size of a sensor pixel in the layer being 
                  considered, and $\beta$ is the number of pixels added to the cone radius. 
                  Different values of $\alpha$ 
                  and $\beta$ are used for the ECAL and HCAL with the default values set to
                  $\{\tan\alpha_{\mathrm{E}} = 0.3, \beta_{\mathrm{E}} =1.5\}$, 
                  and $\{\tan\alpha_{\mathrm{H}} = 0.5, 
                  \beta_{\mathrm{H}}=2.5\}$ respectively. Associations may be made with
                  hits in the previous 3 layers. If no association is made, the hit is used to 
                  seed a new cluster. This procedure is repeated sequentially for the hits in each 
                   pseudo-layer (working outward from ECAL front-face).

\smallskip
\noindent
{\bf{iv) Topological Cluster Merging:}} by design the initial clustering errs on the side of splitting up 
      true clusters rather than clustering energy deposits from more than one particle. 
      The next stage of the algorithm is to merge clusters from tracks and hadronic 
      showers which show clear topological signatures of being associated. 
      A number of track-like and shower-like topologies are searched for including looping 
      minimum ionising tracks, back-scattered tracks and showers associated with a 
      hadronic interaction. Before clusters are merged, a simple cut-based photon 
      identification procedure is applied. The cluster merging algorithms are only applied 
      to clusters which have not been identified as photons. 

\smallskip
\noindent
{\bf{v) Statistical Re-clustering:}} The previous four stages of the algorithm were found to 
      perform well 
      for 50\,GeV jets. However, at higher energies the performance degrades rapidly due to
      the increasing overlap between hadronic showers from different particles. To address 
      this, temporary associations of tracks with reconstructed calorimeter clusters are made. 
      If the track momentum is incompatible with the energy of the associated 
      cluster re-clustering is performed. If $E_{\mathrm{CAL}}-E_{\mathrm{TRACK}} > 3.5\sigma_E$, 
      where $\sigma_E$ is the energy resolution of the cluster, 
      the clustering algorithm, described in {\em iii)} and {\em iv)} above, 
      is reapplied to the hits in that cluster.
      This is repeated, using successively smaller values of the $\alpha$s and $\beta$s in the 
      clustering finding algorithm (stage {\em iii)}) until the 
      cluster splits to give an acceptable track-cluster energy match. Similarly,
      if $E_{\mathrm{TRACK}}-E_{\mathrm{CAL}} > 3.5\sigma_E$ the algorithm attempts 
      to merge additional clusters with the cluster associated with the track. In doing
      so high energy clusters may be split as above.

\smallskip
\noindent
{\bf{vi) Formation of Particle Flow Objects:}} 
     The final stage of the algorithm is to create Particle Flow Objects (PFOs) from
     the results of the clustering. Tracks are matched to clusters on the basis 
     of the distance closest approach of the track projection into the first 10 layers 
     of the calorimeter. If a hit is found within 50\,mm of the track extrapolation an 
     association is made. The reconstructed PFOs are written out in \LCIO\cite{bib:marlin} 
     format.
 
\section{Performance}

\begin{center}
\begin{figure*}[t]
\epsfxsize=15.5cm
\centerline{\epsfbox{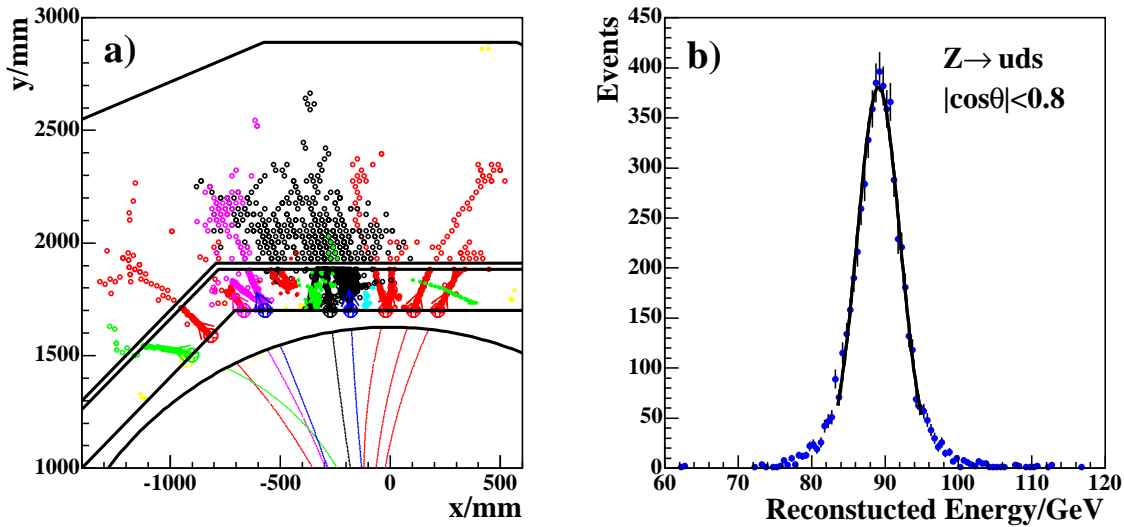}}
\caption{a) \PANDORAPFA\ reconstruction of a 100\,GeV jet in the \MOKKA\
             simulation of the
             Tesla TDR detector. b) The total reconstructed energy from reconstructed 
             PFOs in $\Zzero\rightarrow{\mathrm{uds}}$ events for initial quark directions
             within the polar angle acceptance $|\cosqq|<0.8$. The solid line shows a 
             Gaussian fit to the peak region with a standard deviation of 2.9\,GeV.}
\label{fig:figure1}
\end{figure*}
\end{center}

Fig.~\ref{fig:figure1}a) shows an example of a \PANDORAPFA\ reconstruction of 
a 100\,GeV jet from a $\Zzero\rightarrow\uu$ decay at $\roots=200$\,GeV. The ability
to track particles in the high granularity Tesla TDR calorimeter can be seen clearly. 
Fig.~1b) shows the total PFA reconstructed 
energy for $\Zzero\rightarrow{\mathrm{uds}}$ events with $|\cosqq|<0.8$, 
where $\thetaqq$ is the polar angle of the generated $\qq$ system. These events were
generated at $\roots=91.2$\,GeV using the Tesla TDR detector model.
The root-mean-square deviation from the mean (rms) 
of the distribution is 4.0\,GeV. However, quoting the rms as a measure of the performance 
over-emphasises the importance of the tails. For example, in this figure, the central peak 
is well described by a Gaussian of width 2.9\,\GeV, equivalent to a resolution of 
$\sigma_E/E = 0.31/\sqrt{\EGeV}$. In this paper two measures of the performance are quoted.
The first measure, $\rms$, is the rms in the smallest range of reconstructed energy which 
contains 90\,\% of the events. The second performance measure is obtained from a fit to 
the reconstructed energy distribution. The fit function is the sum of two Gaussian distributions
with a common mean but different widths. The width of the narrower Gaussian, which is constrained 
to contain 75\,\% of the events, gives a measure of the resolution in the peak, $\sigma_{75}$.  
For the data shown in Fig.~\ref{fig:figure1}b) both methods give a resolution of
$\sigma_E/E = 0.3/\sqrt{\EGeV}$;  the ILC goal. However, this is of little consequence to 
ILC physics where, in general, the jets will be higher in energy. 

The majority of interesting ILC physics will consist of final states with
at least six fermions, setting a ``typical'' energy scale for ILC jets 
as approximately 85\,GeV and 170\,GeV at $\roots=500$\,GeV and $\roots=$1\,TeV respectively. 
Fig.~\ref{fig:figure2}a shows the reconstructed total energy in $\Zzero\rightarrow$uds
events (generated without ISR or beamstrahlung effects) at $\roots=360$\,GeV. The fit 
to the sum of a double Gaussian gives $\sigma_{75}=10.8$\,GeV, equivalent to a resolution of
$\sigma_E/E = 0.57/\sqrt{\EGeV}$, significantly worse than that obtained for lower energy jets.
Fig.~\ref{fig:figure2} shows the jet energy resolution for $\Zzero\rightarrow$uds events 
plotted against $|\cosqq|$ for four different values of $\roots$.

\section{Discussion}
\begin{center}
\begin{figure*}[tp]
\epsfxsize=15.5cm
\centerline{\epsfbox{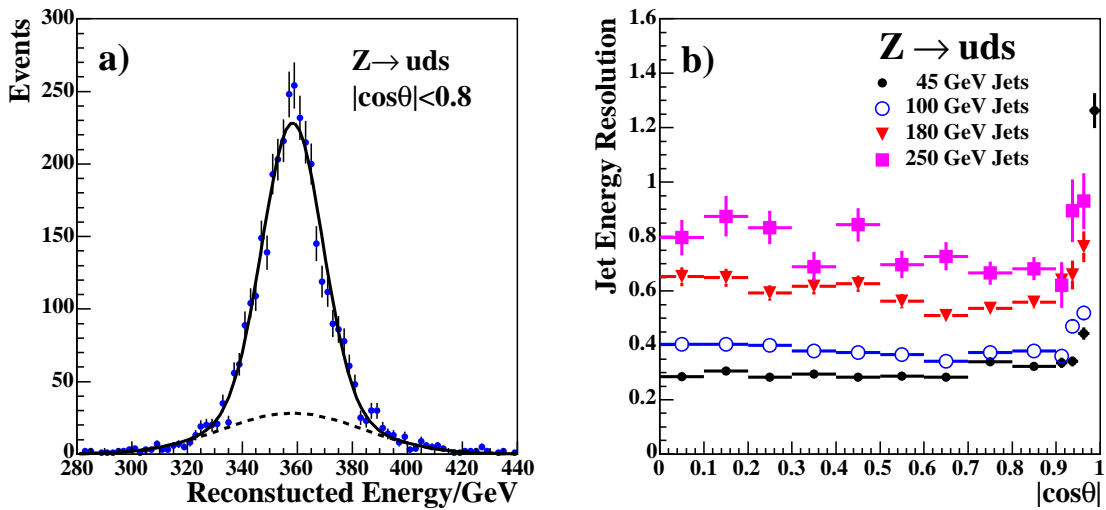}}
\caption{a) The total reconstructed energy from reconstructed 
             PFOs in $\Zzero\rightarrow{\mathrm{uds}}$ at $\roots=360$\,GeV for initial 
             quark directions within the polar angle acceptance $|\cos\theta|<0.8$. The solid line 
             shows a results of the fit to two Gaussians and the dashed line indicates the contribution
             from the broader Gaussian which is constrained to contain 25\,\% of the events.
         b) The jet energy resolution, defined as the $\alpha$ in $\sigma_E/E=\alpha\sqrt{\EGeV}$, 
             plotted versus $\cosqq$ for four different values of $\roots$.}
\label{fig:figure2}
\end{figure*}
\end{center}

The results described above are summarised in Tab.~\ref{tab:resvsE}. The observed jet 
energy resolution in simulated events is not described by the expression 
$\sigma_E/E = \alpha/\sqrt{\EGeV}$. This is not surprising, as the particle density increases it becomes harder to correctly
associate the calorimetric energy deposits to the particles and the confusion term increases.
Empirically it is found that the total energy resolutions in Tab.~\ref{tab:resvsE} can be
described by a {\em jet} energy resolution of 
$\sigma_E/E = 0.265/\sqrt{\EGeV} + 1.2\times10^{-4}\EGeV$, where $E$ is the {\em jet} energy.
This expression represents the current performance of the \PANDORAPFA\ algorithm and should not
be be considered as anything more fundamental. It should be noted that in the current 
\MOKKA\  simulation of the Tesla TDR detector the muon chambers are not included. In principle 
these can be used as a ``tail-catcher'' to improve the energy measurement for high 
energy hadronic showers which may not be fully contained in the HCAL. In the current version 
of \PANDORAPFA\ no attempt is made to correct for this energy leakage. It is noticeable in 
Fig.~\ref{fig:figure2}b that the energy resolution improves with increasing
polar angle in the barrel region of the detector, possibly due to increasing shower containment.
\begin{table*}[bth]
\begin{tabular}{rrcrc}
  Jet Energy        & $\rms$ &  $\rms/\sqrt{\EGeV}$ & $\sigma_{75}$ &$\sigma_{75}/\sqrt{\EGeV}$ \\ \hline
  45 GeV            & 2.8\,GeV        &  0.30            & 2.8\,GeV &  0.30  \\
  100 GeV           & 5.3\,GeV        &  0.38            & 5.2\,GeV &  0.37  \\
  180 GeV           & 11.0\,GeV       &  0.58            &10.8\,GeV &  0.57  \\
  250 GeV           & 16.8\,GeV       &  0.76            &16.8\,GeV &  0.75  \\
\end{tabular}
\caption{Jet energy resolution, expressed as both $\rms$ and $\sigma_{75}$, 
    for $\Zzero\rightarrow$uds events with $|\cosqq|<0.8$.
\label{tab:resvsE}}
\end{table*}

\section{Conclusions}  

Particle flow calorimetry is widely believed to be the key to reaching the ILC jet 
energy resolution goal of $\sigma_E/E = 0.3/\sqrt{\EGeV}$. Consequently, the design 
and optimisation of detectors for the ILC depends both on hardware and on sophisticated
software reconstruction. For the Tesla TDR detector concept, the \PANDORAPFA\ particle 
flow algorithm achieves good performance, $<0.4/\sqrt{\EGeV}$, for jet energies upto 
about 100\,\GeV. For higher energies the performance degrades significantly 
reaching the equivalent of $0.6/\sqrt{\EGeV}$ for 200\,GeV jets. With further 
optimisation of the algorithm the performance is expected to improve. However, the
current algorithm is adequate for most ILC physics studies at $\roots=500$\,GeV and
may be used for the optimisation of the design of the ILC detector(s).

\end{document}